\documentclass[twocolumn,showpacs,aps,prl,superscriptaddress]{revtex4}

\usepackage{graphicx}
\usepackage{amsmath}

\begin{document}

\title{Linear temperature dependence of conductivity in Si
two-dimensional electrons near the apparent metal-to-insulator
transition}

\author{K. Lai$^\ast$}
\affiliation{Department of Electrical Engineering, Princeton
University, Princeton, New Jersey 08544}
\author{W. Pan}
\affiliation{Sandia National Laboratories, Albuquerque, NM 87185}
\author{D.C. Tsui} \affiliation{Department of Electrical
Engineering, Princeton University, Princeton, New Jersey 08544}
\author{S. Lyon}
\affiliation{Department of Electrical Engineering, Princeton
University, Princeton, New Jersey 08544}
\author{M. M\"uhlberger}
\affiliation{Institut f\"ur Halbleiterphysik, Universit\"at Linz,
A-4040 Linz, Austria}
\author{F. Sch\"affler}
\affiliation{Institut f\"ur Halbleiterphysik, Universit\"at Linz,
A-4040 Linz, Austria}

\date{\today}

\begin{abstract}

In a high mobility two-dimensional electron system in Si, near the
critical density, $n_c=0.32\times10^{11}$cm$^{-2}$, of the apparent
metal-to-insulator transition, the conductivity displays a linear
temperature ($T$) dependence around the Fermi temperature. When
$\sigma_0$, the extrapolated $T=0$ conductivity from the linear
T-dependence, is plotted as a function of density, two regimes with
different $\sigma_0(n)$ relations are seen, suggestive of two
different phases. Interestingly, a sharp transition between these
two regimes coincides with $n_c$, and $\sigma_0$ of the transition
is $\sim$ $e^2/h$, the quantum conductance, per square. Toward
$T=0$, the data deviate from linear $\sigma(T)$ relation and we
discuss the possible percolation type of transition in our Si
sample.

\end{abstract}
\maketitle

The nature of the ground state of an interacting two-dimensional
electron system (2DES) in the presence of disorder is a
long-standing problem in condensed matter physics. More than a
decade ago, an apparent 2D metal-to-insulator transition (MIT) was
first reported as the density of a Si MOSFET is reduced through a
characteristic density $n_c$ \cite{krav1}. Despite much research
effort later in the field \cite{abrahams, krav2}, there remain
several unsettled fundamental questions. For example, metallic
behavior is observed for density $n>n_c$. Does this metallic-like
state persist down to $T=0$ and thus represent a true 2D metal?
Besides, the 2DES eventually becomes insulating as ever-decreasing
densities. Is this phenomenon related to a novel phase transition
due to strong electron-electron (e-e) interactions \cite{krav2}, or
is it a mundane crossover just due to a complex combination of many
well-understood physical mechanisms \cite{das1}?

In an earlier publication, we reported the observation of the
metallic behavior and the apparent 2D MIT in a high mobility Si
quantum well \cite{lai1}. The high-density metallic-like state and
its response to an in-plane magnetic field are emphasized in that
paper. In this communication, we focus on the transport properties
for densities close to the transition. Here, the conductivity
displays a linear temperature dependence near $T=T_F$ (the Fermi
temperature) and the slope is the same for all different densities
around $n_c$. When the extrapolated $T=0$ conductivity, $\sigma_0$,
of this T-dependence is plotted as a function of density, two
regimes with linear $\sigma_0$ vs $n$ relations are readily seen.
Interestingly, these two $\sigma_0(n)$ lines cross each other
exactly at $n_c$ and $\sigma_0$ at the crossing coincides with the
quantum conductance, $e^2/h$, per square. At low temperatures,
$T<<T_F$, the measured $\sigma(T)$ deviates from the linear
T-dependence. We discuss the low-T behavior of our data within the
percolation model.

The experiments were performed on the 2DES in an n-type Si quantum
well confined in a Si$_{0.75}$Ge$_{0.25}$/Si/Si$_{0.75}$Ge$_{0.25}$
heterostructure. The 2D electron density is tuned continuously by
applying a front gate voltage to our field-effect transistor device.
Details of the growth and the sample structure can be found in Ref.
\cite{lai2}. Standard low-frequency ($\sim$ 7Hz) lock-in techniques
were used to measure the 2D resistivity $\rho$. At $T\sim300$mK and
zero gate voltage, the 2DES has a density
$n=1.45\times10^{11}$cm$^{-2}$ and mobility $\mu=190,000$cm$^2$/Vs.

In Fig. 1(a), we reproduce selected T-dependence data $\rho(T)$ from
our previous paper \cite{lai1}. Toward the $T=0$ limit, the apparent
2D MIT is clearly seen at the critical density
$n_c=0.32\times10^{11}$cm$^{-2}$, where d$\rho$/d$T$ $\sim$ 0 for
$T<1$K. The insulating behavior, d$\rho$/d$T<0$, is observed for
densities below $n_c$ and metallic-like behavior, d$\rho$/d$T>0$,
above $n_c$. For further insights to the 2D MIT, we now focus on the
data around $n_c$ and plot the inverse resistivity, or conductivity
$\sigma$, as a function of temperature in the density range of
$0.27\times10^{11}<n<0.38\times10^{11}$cm$^{-2}$ in Fig. 1(b). In
the low-T limit, the MIT is again observed in that d$\sigma$/d$T$
changes sign as $n$ changes through $n_c$. At high temperatures when
$T$ approaches the Fermi temperature $T_F$, marked as short lines
for each density, however, all $\sigma(T)$ curves show roughly a
linear T-dependence and bends slightly downward for $T$ sufficiently
higher than $T_F$.

\begin{figure}[!t]
\begin{center}
\includegraphics[width=3.2in,trim=0.3in 0.4in 0.2in 0.2in]{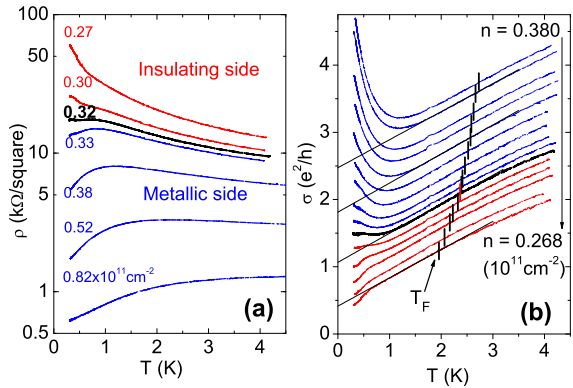}
\end{center}
\caption{\label{1} (a) Selected data of $\rho(T)$ in a high quality
Si quantum well reproduced from Fig. 1 in Ref. [5]. (b) $\sigma(T)$
in the vicinity of the critical density. The electron densities,
from bottom to top, are 0.268, 0.282, 0.296, 0.303, 0.318, 0.322,
0.324, 0.328, 0.336, 0.344, 0.352, 0.358, 0.365, 0.374, and 0.380,
in units of 10$^{11}$cm$^{-2}$, respectively. The short, vertical
lines mark the Fermi energy at each density. The lines on some
curves are the linear fits around $T=T_F$.}
\end{figure}

To illustrate this linear $\sigma(T)$ relation around $T=T_F$, we
show the linear fits, $\sigma(T)=\sigma_0+\gamma$$T$, for several
densities, where $\gamma$ is the slope and $\sigma_0$ the linear
extrapolation to $T=0$. In the inset of Fig. 2, $\gamma$ is plotted
as a function of the density. It is nearly constant,
$\sim0.43\pm0.01e^2/h$ per Kelvin, for $n<0.38\times
10^{11}$cm$^{-2}$ and then rapidly decreases for higher densities.
We plot $\sigma_0$ vs $n$ in Fig. 2 and two regimes with different
linear $\sigma_0(n)$ relations are readily identified. On the
low-density insulating side, $\sigma_0$ increases as increasing
density at a rate of 11 $e^2/h$ per 10$^{11}$cm$^{-2}$. The slope
roughly doubles to 26 $e^2/h$ per 10$^{11}$cm$^{-2}$ on the
high-density side. Strikingly, the $\sigma_0(n)$ data show a sharp
bend at the crossing of the two straight lines at
$n=0.32\times10^{11}$cm$^{-2}$, which coincides with $n_c$ of the 2D
MIT. And $\sigma_0$ at this density exactly equals the quantum
conductance, $e^2/h$, per square. For both regimes, we extrapolate
the linear $\sigma_0(n)$ to zero conductivity at
$n_1=0.23\times10^{11}$cm$^{-2}$ on the low-density side and
$n_2=0.28\times10^{11}$cm$^{-2}$ on the high-density side.

\begin{figure}[!t]
\begin{center}
\includegraphics[width=3.2in,trim=0.3in 0.5in 0.2in 0.2in]{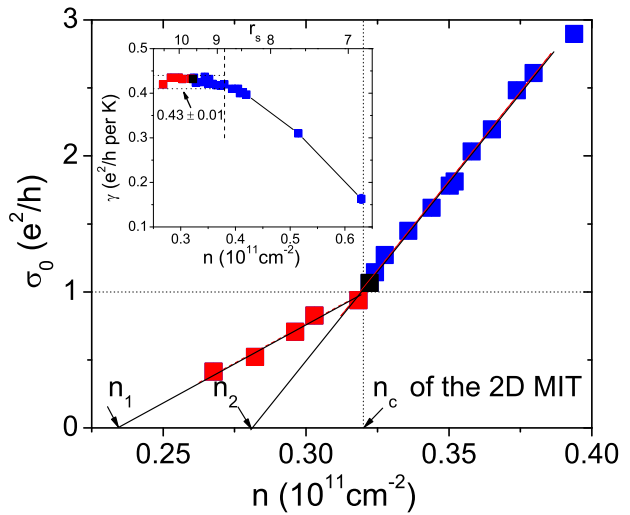}
\end{center}
\caption{\label{2} The $T=0$ conductivity, $\sigma_0$, extrapolated
from the high-T linear fit in Fig. 1(b), is plotted as a function of
the 2DES density. The solid lines are linear fits on both low-n and
high-n regimes and they extrapolate to $\sigma_0=0$ at
$n_1=0.23\times10^{11}$cm$^{-2}$, and
$n_2=0.28\times10^{11}$cm$^{-2}$, respectively. The dotted lines
show that the two linear fits cross at $n_c$ and $\sigma_0=e^2/h$.
The inset shows the slope, $\gamma$, of the linear $\sigma(T)$ in
Fig. 1(b), as a function of density. is roughly constant below
$0.38\times10^{11}$cm$^{-2}$, and decreases rapidly for higher
densities.}
\end{figure}

Linear $\sigma(T)$ relation has been observed before in p-type GaAs
samples \cite{noh, huang}. There, $\gamma$ = d$\sigma$/d$T$ $\sim$ 3
to 5 $e^2/h$ per Kelvin, compared to the much smaller number,
$\sim0.43e^2/h$ per Kelvin, in our Si quantum well sample. This one
order of magnitude difference might be related to the different
Fermi temperatures in these two systems. As will be shown below,
d$\sigma$/d$T$ $\sim$ 1/$T_F$. In our Si sample, $T_F$ ($\sim$ 2 to
2.5K) is about a factor of 10 larger than that ($\sim$ 0.1 to 0.4 K)
in the p-GaAs samples. As a result, d$\sigma$/d$T$ in Si is expected
to be $\sim$10 times smaller than in p-GaAs.

Theoretically, the linear $\sigma(T)$ in the high temperature regime
has also been addressed by several models. Das Sarma and Hwang
\cite{das2} calculated the temperature dependence of $\sigma$ within
a classical model of the screened charged impurity scattering. For
$T>>T_F$, they observed that $\sigma(T)\sim$ $T/T_F$ $\propto$ $T$,
consistent with our experimental observation. Besides the screening
model, the micro-emulsion model proposed by Spivak and Kivelson
\cite{spivak} can also explain the linear ($T$) at high
temperatures. In this model, near $n_c$, the ground state of the
2DES consists of an electron liquid with Wigner crystal inclusions
on the metallic side, and Wigner solid with liquid inclusions on the
insulating side. At high temperatures, Wigner crystal droplets melt
and the system behaves classically. In analogy to the physics of
$^3$He near the crystallization pressure \cite{spivak}, the
viscosity of the electron liquid, which is directly proportional to
the resistivity, is inversely proportional to $T$, resulting in a
linear $\sigma(T)$, again consistent with our observation
\cite{novikov}.

Quantitatively, however, neither of the two above models can explain
the experimental observation that d$\sigma$/d$T$ is nearly
independent of the electron density around $n_c$. Under the
temperature dependent screening model, d$\sigma$/d$T$ is
proportional to $1/T_F$ or $n^{-1}$ \cite{das2}. As a result, the
slope should decrease by a factor of $\sim$ 1.5 when the 2DES
density increases from 0.27 to $0.38\times10^{11}$cm$^{-2}$. On the
other hand, the micro-emulsion model shows d$\sigma$/d$T$ $\sim$
$n^2$ \cite{spivak}. Consequently, in the same density range, the
slope should change roughly by a factor of 2. In contrast,
d$\sigma$/d$T$ is nearly constant and $\sim0.43\pm0.01e^2/h$ per
Kelvin in our measurements. So far, it is not known what is
responsible for this inconsistency between our experimental result
and the theoretical predictions.

Having discussed the linear T-dependence of the data, we need to
address some puzzling aspects of the overall results. First, the
examined specimen has very high electron mobility and the $n_c$ of
the 2D MIT is by far the lowest among all the Si-based samples. The
e-e interaction parameter $r_s$ at the transition density is $\sim$
10, i.e., the Coulomb energy $E_c$ exceeds the Fermi energy $E_F$ by
a factor of 20, after taking into account the two-fold valley
degeneracy in (001) Si 2DES. Consequently, even though the system
behaves classically for $T$ $\sim$ $T_F$, the 2DES is still strongly
correlated since $E_c$ is the dominant energy scale here. Second, if
the high-T physics underlying the linear $\sigma(T)$ behavior
persists to the $T=0$ limit, our $\sigma_0(n)$ data show two
distinct regimes, possibly suggesting two different electronic
phases below and above $n_c$.  The transition between the two
density regimes is sharp, and occurs almost exactly at $n_c$.
$\sigma_0$ at this transition point is very close to $e^2/h$, the
unit of quantum conductance, per square, manifesting a possible
quantum nature of this transition. The reduction of the rate,
d$\sigma_0$/d$n$, of the density dependence, from $\sim26e^2/h$ per
10$^{11}$ cm$^{-2}$ above $n_c$ to $\sim11e^2/h$ per 10$^{11}$
cm$^{-2}$ below $n_c$, clearly indicates that in the low-density
regime the 2D electrons are less likely to become localized by
reducing the 2DES density. Considering the large $r_s$ at $n_c$, one
might speculate that strong e-e interactions help to prevent
electrons from being localized at low densities. Indeed, in a recent
publication \cite{shi}, Shi and Xie showed that, with e-e
interactions taken into account, the 2DES becomes less localized
compared to a non-interacting system at the same density. However,
the appearance of two density regimes is still unexplained because a
smooth evolution should be expected from their calculations.

\begin{figure}[!t]
\begin{center}
\includegraphics[width=3.2in,trim=0.3in 0.5in 0.2in 0.2in]{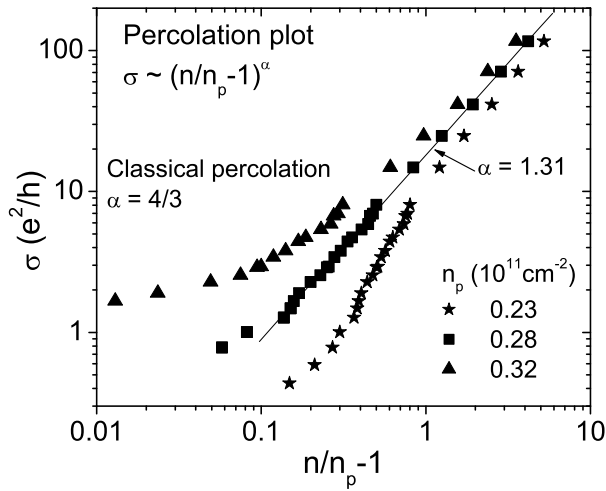}
\end{center}
\caption{\label{3} Percolation plot of $\sigma$(0.3K) vs $n/n_p-1$
in a log-log scale. Three trial densities at 0.23, 0.28, and
0.32$\times10^{11}$cm$^{-2}$, indicated by different symbols, are
used as $n_p$ in the plot. For $n_p=0.28\times10^{11}$cm$^{-2}$,
excluding the two lowest density points, a linear fit (solid line)
is obtained over two decades in $\sigma$. And the slope corresponds
to an exponent of $\alpha=1.31$.}
\end{figure}

We now turn to the low-T limit of the measured $\sigma(T)$ in Fig.
1(b). Instead of following the trend from high temperatures, the
conductivity deviates from the linear T-dependence toward the $T=0$
ground state configuration. It has long been suggested that the
apparent MIT observed here might be of a density inhomogeneity
effect, and belongs to a general class of 2D percolation problem
\cite{efros, he}. In this picture, at very low densities, the 2DES
is macroscopically inhomogeneous and first forms isolated puddles by
occupying the low potential ``valleys''. As $n$ increases, the area
of the electron puddles increases and at the percolation threshold
density $n_p$, some puddles are connected, giving rise to a
conducting path throughout the sample. Experimentally, this
percolation-type 2D MIT is supported by various measurements, e.g.,
scanning near-field optical microscopy (SNOM) \cite{eytan}, scanning
single-electron transistor (SET) microscopy \cite{ilani}, transport
\cite{das3}, and surface-acoustic wave (SAW) \cite{tracy}
experiments.

According to the percolation model, the conductivity of the 2DES
follows the scaling function $\sigma\sim(n/n_p-1)^\alpha$, and the
exponent $\alpha$ in a classical percolation transition is 4/3 [14].
In Fig. 3, we show the measured $\sigma$ at $T=0.3$K as a function
of $n/n_p-1$ in a log-log scale. Since $n_p$ is not known beforehand
from finite temperature measurements, three trial densities of $n_p$
are used, in unit of 10$^{11}$cm$^{-2}$, $n_1=0.23$ (extrapolation
to $\sigma_0=0$ from the low-n regime), $n_2=0.28$ (extrapolation to
$\sigma_0=0$ from the high-n regime), and $n_c=0.32$ (critical
density of the observed 2D MIT). For $n_p=n_2=0.28\times10^{11}$
cm$^{-2}$, except for some deviation at very low densities, a good
linear fit with $\alpha=1.31$, close to 4/3 in the classical model,
can be obtained over two decades in $\sigma$. It is not known, at
this stage, whether this coincidence is accidental or the two
densities are actually deeply related. On the other hand, for
$n_p=n_1$ or $n_c$, the fittings to a power law behavior are poor.
We nevertheless emphasize two important points when applying the
percolation model to our data. First, the power law of conductivity
in the percolation model only holds for densities very close to
$n_p$, or $n/n_p-1<<1$ [13]. Second, percolation transition is
essentially a zero-temperature phase transition. As seen in Fig.
1(b), $\sigma(T)$ does not saturate at our lowest measured $T=0.3$K.
In this regard, it is necessary that further measurements be carried
out at lower temperatures.

In summary, in a high quality Si quantum well specimen, near the
apparent 2D metal-to-insulator transition, a linear temperature
dependence of conductivity is observed at $T$ around $T_F$ on both
sides of $n_c$. When $\sigma_0$, the extrapolation of this linear
$\sigma(T)$ to $T=0$, is plotted as a function of density, two
regimes with different $\sigma_0$ vs $n$ relations are readily seen.
Interestingly, the two linear $\sigma_0(n)$ regions cross almost
exactly at $n_c$, and $\sigma_0$ at the crossing point is $e^2/h$,
the quantum conductance, per square. We also show that the measured
$\sigma(T)$ at our low-T limit can be fitted by a percolation
scaling function $\sigma\sim(n/n_p-1)^\alpha$ when $n_p=n_2$, the
extrapolation of $\sigma_0(n)$ to $T=0$ on the high-density side.

The work at Princeton was supported by AFOSR under grant No.
0190-G-FB463 and the NSF DMR-0352-533 and DMR-02-13706. The part of
work carried out in Linz was supported by Gme and FWF
(P-162223-N08), both Vienna, Austria. Sandia National Laboratories
is a multiprogram laboratory operated by Sandia Corporation, a
Lockheed-Martin company, for the U.S. Department of Energy under
Contract No. DE-AC04-94AL85000. We thank R. Bhatt and D. Novikov for
illuminating discussions.
\\
\\
\noindent $^\ast$ Present address: Department of Applied Physics,
Stanford University, Stanford, California 94305.


\begin{thebibliography}{90}
\bibitem{novikov} We note that the viscosity of the electron liquid
was also calculated by D.S. Novikov, cond-mat/0603184, and is shown
to be non-analytical in the relevant temperature range.
\end{thebibliography}
\end{document}